\documentclass[preprint]{aastex}
\usepackage{amsmath}
\usepackage{cases}
\usepackage{url}
\usepackage{color}
\usepackage{lineno}

\shorttitle{Fast Reconnection by Turbulence}
\shortauthors{Yang et al.}

\begin{document}

\title{Fast Magnetic Reconnection with Turbulence in High Lundquist Number Limit}

\author{Liping Yang\altaffilmark{1,2}, Hui Li\altaffilmark{2}, Fan Guo\altaffilmark{2}, Xiaocan Li\altaffilmark{2}, Shengtai Li\altaffilmark{2}, Jiansen He\altaffilmark{2}, Lei Zhang\altaffilmark{1}, Xueshang Feng\altaffilmark{1}}

\altaffiltext{1}{SIGMA Weather Group, State Key Laboratory for Space
Weather, National Space Science Center, Chinese
Academy of Sciences, 100190, Beijing, China, lpyang@swl.ac.cn}
\altaffiltext{2}{Theoretical Division, MS B284, Los Alamos National Laboratory, Los Alamos, NM 87545, USA}
\altaffiltext{3}{School of Earth and Space Sciences, Peking University, 100871 Beijing,
China}

\begin{abstract}
We use extensive 3D resistive MHD simulations
to study how large-scale current sheets will undergo fast reconnection in
the high Lundquist number $S$ limit (above $\sim 10^4$), when the system is subject
to different externally driven turbulence levels and the self-generated
turbulence produced by 3D reconnection dynamics.
We find that the normalized global reconnection rate
$\sim 0.01 - 0.13$, weakly dependent on $S$.
Global reconnection
with the classic inflow/outflow configurations is observed, and 3D flux ropes
are hierarchically formed and ejected from reconnection regions.
A statistical separation of the reconnected magnetic field lines follows a super-diffusive
behavior, from which the rate is measured to be very similar to
that obtained from the mixing of tracer populations.
We find that the reconnection rate scales roughly linearly with the
turbulence level during the peak of reconnection. This scaling is consistent
with the turbulence properties produced by both the
externally driven and self-generation processes. These results
imply that large-scale thin current sheets tend to undergo rigorous reconnection.

\end{abstract}


\section{INTRODUCTION}

Fast magnetic reconnection (at a fraction of Alfv\'en speed $V_A$)
is often invoked to explain
energetic events such as
solar/stellar  flares,  substorms in the magnetosphere of
Earth and other planets,  coronal mass ejections,
sawtooth crashes in fusion plasmas, and
other astrophysical systems \citep{Priest2007, Lazarian2020}.
During reconnection,
oppositely directed magnetic field lines restructure themselves,
resulting in a rapid conversion of magnetic energy into kinetic energy of bulk flows, and thermal
and non-thermal particles \citep[e.g.,][]{Drake2006b}.

In the limit of  resistive magnetohydrodynamics (MHD) description,
the classical Sweet-Parker (SP) model \citep{Sweet1958, Parker1957} predicts a rather slow
reconnection rate
proportional to $S^{-1/2}$, where $S = L V_A / \eta$ is the Lundquist number,
$\eta$ is the plasma resistivity, and $L$ is the characteristic length of the system. Many alternatives to speed up
the reconnection have been investigated
 \citep{Priest2007, Cassak2012, Lin2015, Loureiro2016}.
A major advance came through studies related to the resistive tearing instability
\citep{Biskamp1986, Loureiro2007, Bhattacharjee2009,
Uzdensky2010, Huang2010, Ni2012, Lin2018}.
In the high $S$ limit,  it is found that,
above a critical $S \sim 10^4$, the thin SP current sheets (CSs) in two-dimensional (2D)
become violently unstable to the hierarchical formation and ejection
of plasmoids \citep{Loureiro2007}, producing nearly resistivity-independent
reconnection rate around $0.01 \ V_A$.

Fast magnetic reconnection in the presence of 3D turbulence is a critically
important process  in space and astrophysical plasmas
\citep{Matthaeus1986, Lazarian1999, Fan2004, Kowal2009,
Loureiro2009, Eyink2011, Daughton2011,
Wyper2015, Oishi2015, Takamoto2015, Guo2015,
Huang2016, Beresnyak2017, Kowal2017,
Pisokas2018, Li_X2019, Ye2020}, with some interesting
observation support \citep{Fu2017,  He2018, Chitta2020}.
Broadly speaking three types of configurations have been studied in
some detail, depending on what ``free energy" is available.
The first is on how externally driven (or decaying) turbulence
affects the reconnection of a pre-existing CS(s)
\citep[e.g.,][]{Matthaeus1986, Lazarian1999,  Kowal2009, Loureiro2009, Kowal2012}. The 3D MHD simulations have
mostly been done in the small $S$ ($\sim$ a few thousand) limit, though the externally
driven turbulence  with relatively large amplitude
can greatly enhance the reconnection rate up to $\sim 0.1\ V_A$.
The second is similar to the first type except that the turbulence is self-generated
from instabilities associated with the pre-existing CS(s)
or additional instabilities due to reconnection
\citep[e.g.,][]{Oishi2015, Huang2016, Beresnyak2017, Kowal2017, Kowal2020}.
The 3D MHD simulations in this category with $S$ up to a few times $10^5$
have shown that the reconnection rate is slightly slower,
averaging around a few percent of $V_A$.
Note that these two types of studies could differ in important ways
because the available free energy in the second case is primarily
from the initial CS only
whereas in the first case both the injected turbulence and the
CS contribute to the available energy for dissipation.
In particular, \citet[][hereafter LV99]{Lazarian1999} and
\citet[][hereafter ELV11]{Eyink2011} provided the basic theoretical model
on such turbulent reconnection.
The third is to begin with the injected turbulence only without a
pre-existing semi-global CS(s). The turbulence cascade
will produce CSs at intermediate scales that could undergo
reconnection. 2D MHD simulations
\citep{Dong2018, Walker2018} and 3D kinetic simulations
\citep[e.g.,][]{Makwana2015} appear to lend support to these ideas.
In fact, the dual process of CS formation by turbulence cascade and
the back-reaction on turbulence by the possible reconnection of such sheets
have led to new models of MHD turbulence with reconnection \citep[e.g.,][]{Loureiro2017, Boldyrev2017}.
Note that the available free energy in this case is only the injected
turbulence, very different from the first two types.
Overall, the interplay among the externally injected turbulence vs.
the self-generated turbulence, and the pre-existing CS(s)
vs. the self-generated CSs makes it challenging to
build a comprehensive theory. Numerical simulations tend to have a
limited dynamic range to fully resolve several critical issues
revealed by these theoretical models (see a recent
discussion in \citet{Lazarian2020}).

In this work, we use a set of 3D compressible MHD simulations
to systematically examine how the
reconnection rate  in the low plasma $\beta = 0.1$ condition
scales with the strength of turbulence as well as $S$.
Our most important conclusion is that, in systems with an initial large-scale CS,
the 3D reconnection rate can range between $0.01 - 0.1 \ V_A$,
and scales roughly linearly with the turbulent  Alfv\'en Mach number
$M_{\mathrm{A}} \sim 0.06 - 0.32$.
The rate is weakly dependent on $S$ in the high $S$ limit.
Flux ropes, as the 3D version of the 2D plasmoid instability,
are frequently formed and ejected along the thin CSs.
Magnetic field line tracing yields super-diffusive behavior.
The turbulence is a combination of the externally driven
and the self-generated fluctuations, but with a second-order structure
function different from the incompressible MHD turbulence
theory by \citet{Goldreich1995}.

\section{NUMERICAL MHD MODEL}
\label{sec:model}
The isothermal resistive
MHD equations in a periodic cube with a
side length of $L = 2 \pi$ are solved:
\begin{equation}
 \frac{\partial \rho}{\partial
t}+ \nabla \cdot (\rho \mathbf{u}) = 0 \ ,
\end{equation}
\begin{equation}
 \frac{\partial \rho \mathbf{u}}{\partial
t}+ \nabla \cdot \left[\rho \mathbf{u} \mathbf{u} + ( p +
\frac{1}{2}\mathbf{B}^2 )\mathbf{I}-\mathbf{B} \mathbf{B}\right] = \nu\nabla^2
\mathbf{u} + \rho\mathbf{f}_v\ ,
\end{equation}
 \begin{equation}
\frac{\partial \mathbf{B}}{\partial t}+ \nabla \cdot
(\mathbf{u}\mathbf{B}-\mathbf{B}\mathbf{u}) = \eta\nabla^2
\mathbf{B} \ .
\end{equation}
\begin{equation}
 \frac{\partial s_i}{\partial
t}+ \nabla \cdot (s_i \mathbf{u}) = 0 \ ,
\end{equation}
Here, $\rho$ is the mass density; $p$ is the thermal pressure; $\mathbf{u}$
is the velocity; $\mathbf{B}$
denotes the magnetic field; $t$ is time; $\nu$ is the viscosity;
$\eta$ is the magnetic resistivity; $s_i (i=1,2) $ are the densities
of the tracer populations \citep{Yang2013a};
$\mathbf{f}_v$ is a random large-scale driving force,
applied in Fourier space at $ k < 3.5$ \citep{Yang2017a,Yang2018}.
We have used $\nu = \eta$ in all simulations.

The initial magnetic field has a Harris configuration with two thin CSs as\\
$\mathbf{B} = B_0 \left[\tanh(\frac{x-x_1}{w}) -
\tanh(\frac{x-x_2}{w})\right] \hat{y}-B_0\hat{y}$, where
$B_0$ is the asymptotic magnetic field, $x_1=\pi/2$ and $x_2=3\pi/2$ are
the initial positions of the CSs, and the parameter $w$ is set to
satisfy the SP scaling of $2w/L\simeq S^{-1/2}$. Initially,
the density profile is set to maintain a uniform total (thermal plus magnetic)
pressure, velocity is zero, and plasma $\beta$ is about 0.1.
Due to the broadening likely caused by turbulence,
the CS layer during evolution is typically resolved with more than 10 cells.
The externally driven turbulence is characterized by $\mathbf{f}_v$. When
$|\mathbf{f}_v| = 0$, the velocity is initially seeded with a random noise of amplitude  $10^{-3}$.
Simulation parameters are listed in Table 1, in which $N$ is grid number in one direction.
$M_{\mathrm{A}}$ is Alfv\'en Mach number defined as $M_{\mathrm{A}}=u_{\mathrm{RMS}}/V_A $
with $u_{\mathrm{RMS}}$ being the root-mean-square (RMS) amplitude of the velocity
at the peak reconnection, and $V_A$ the Alfv\'en speed based on the
initial magnetic field $B_0$ and the average density.
Run E only has a uniform magnetic field
without any initial CSs.
We use the  Athena code \citep{Gardiner2005,
Stone2008} for simulations. Specifically, we apply the approximate
Riemann solver of
Harten-Lax-van Leer discontinuities (HLLD) to the calculation of the numerical fluxes,
a third-order piecewise parabolic method (PPM) to the reconstruction,
MUSCL-Hancock (VL) Integrator to the time integration, and the
constrained transport (CT) algorithm
to ensure the divergence-free state of the magnetic field.

\section{NUMERICAL RESULTS}
\label{sec:results}

\begin{figure}[htbp]
   \begin{center}
   \begin{tabular}{c}
     \includegraphics[width = 4.5 in]{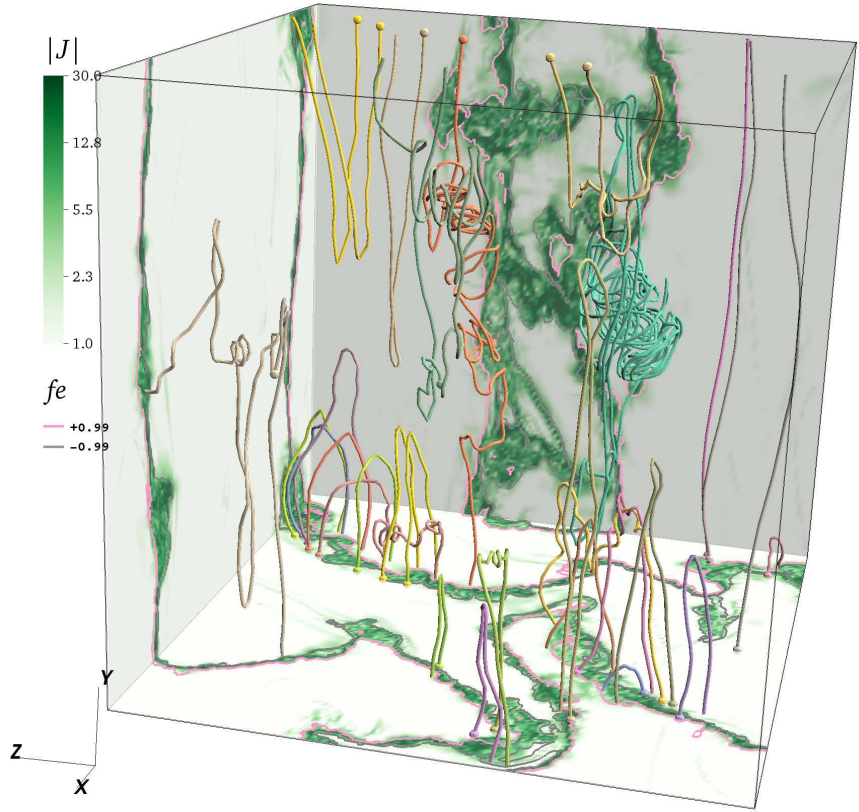}   \\
  \end{tabular}
   \end{center}
\caption{Spatial distribution of the total current density ($|J|$) for Run A1 at
$t = 2.0$. Only intersections with the three bounding planes are shown.
The colored lines denote sample magnetic field lines.
The pink and gray contour lines show $f_e = 0.99$ and $-0.99$, respectively.
}\label{figure1}
\end{figure}

We find that all runs containing initial CSs will undergo
global reconnection characterized by inflow/outflow patterns.
Figure \ref{figure1} shows that the two initially
parallel thin CSs are now strongly deformed
by the externally driven turbulence
while undergoing 3D reconnection. The width of the
CSs also demonstrates
thinning and thickening at various places.
A number of magnetic field lines are plotted and
three typical behaviors are observed: first,
field lines that are relatively smooth and arch-looking
start at one side of a CS and end up at
the other side of the same CS, indicating reconnection accompanied with
$X-$points and large opening angles for reconnected lines;
second, field lines that are far away from the CSs
go through the box without reconnection;
third,  field lines start at one side of a CS but trace out in twisted
trajectories (possibly flux ropes) and end up far away from the starting points.

\begin{table}[ht]

\caption{Reconnection MHD Simulations}
\begin{tabular}{cccccc}
\hline
\hline
Run & $N^3$ & S & $M_{\mathrm{A}}$  &   $|\mathbf{f}_v|$    & CSs \\
\hline
A1 & $2048^3$ & $2.3\times10^5$ & 0.322 & 0.30 & Yes\\
A2 & $1024^3$ & $6.3\times10^4$ & 0.305 & 0.30 & Yes\\
A3 & $1024^3$ & $1.5\times10^4$ & 0.304 & 0.30 & Yes\\
A4 & $1024^3$ &  $4.8\times10^3$ & 0.302 & 0.30 & Yes\\
\hline
B1 & $2048^3$ & $2.3\times10^5$ & 0.192 & 0.10 & Yes\\
B2 & $1024^3$ & $6.3\times10^4$ & 0.185 & 0.10 & Yes\\
B3 & $1024^3$ & $1.5\times10^4$ & 0.183 & 0.10 & Yes\\
B4 & $1024^3$ &  $4.8\times10^3$ & 0.180 & 0.10 & Yes\\
\hline
C1 & $2048^3$ & $2.3\times10^5$ & 0.098 & 0.01 & Yes\\
C2 & $1024^3$ & $6.3\times10^4$ & 0.092 & 0.01 & Yes\\
C3 & $1024^3$ & $1.5\times10^4$ & 0.089 & 0.01 & Yes\\
C4 & $1024^3$ &  $4.8\times10^3$ & 0.084 & 0.01 & Yes\\
\hline
D1 & $2048^3$ & $2.3\times10^5$ & 0.072 & No & Yes\\
D2 & $1024^3$ & $6.3\times10^4$ & 0.067 & No & Yes\\
D3 & $1024^3$ & $1.5\times10^4$ & 0.060 & No & Yes\\
D4 & $1024^3$ &  $4.8\times10^3$ & 0.056 & No & Yes\\
\hline
E & $1024^3$ & $6.3\times10^4$ & 0.421 & 0.30 & No\\
\hline
\end{tabular}
\end{table}

To calculate the 3D reconnection rate, we use the method
described in \citet{Daughton2014}, which employs the mixing of tracer
populations originating from separate sides of a CS as a proxy
to identify the reconnection region and track the evolution of magnetic flux.
We solve Eq. (4) using two tracer species $s_1$ and $s_2$.
The initial values of $s_1$ and  $s_2$ are such that: on one side of a CS, $s_1 = - 1$ and
otherwise $0$, whereas on the other side of the same
CS $s_2 = 1$ and otherwise $0$.
As reconnection proceeds, the populations tagged by $s_1$ and $s_2$
will interpenetrate and  a mixing
fraction $f_e$ can be defined as $f_e =\frac{ |s_1| - |s_2|}{|s_1|+|s_2|}$,
which will  vary continuously from
$f_e = -1$  on one side of CS to $f_e = 1$  to the other side of the same CS.
In Figure \ref{figure1}, we can see that the contours of $f_e$ enclose
the strong $|J|$ layers quite well, correlating strong mixing/reconnection
with strong $|J|$.
The 3D reconnection rate is calculated according to the time
derivative of the unreconnected
magnetic flux $\frac{\partial \Phi}{\partial t}$ within the regions with
$f_e <  - f_c$ or $f_e > f_c$ as $\frac{\partial \Phi}{\partial t}$ is
equal to  the line integral of the
electric field along the surfaces of $f_e = -f_c$ or $f_e = f_c$
due to  the periodic boundary condition.
We have also calculated the change rate of the magnetic flux within the
regions with $f_e \geqslant - f_c$ and $f_e \leqslant f_c$ and found that it is an order
of magnitude smaller than $\frac{\partial \Phi}{\partial t}$. Therefore,
it can be thought that the flux entering into the reconnection region
is dissipated quickly.
Because the boundaries that separate $f_e \ne \pm 1$ regions
from $f_e = \pm 1$ regions are quite sharp, the calculated
reconnection rate is insensitive if $f_c$ is in the range 0.9-0.995
\citep{Daughton2014}. Here, we choose
$f_c = 0.99$. The calculated reconnection rate
grows first as the reconnection starts, reaching a maximum after a few
Alfv\'en times, then gradually decreasing.

\begin{figure}[bp]
   \begin{center}
   \begin{tabular}{c}
     \includegraphics[height=6in, width = 4.5in]{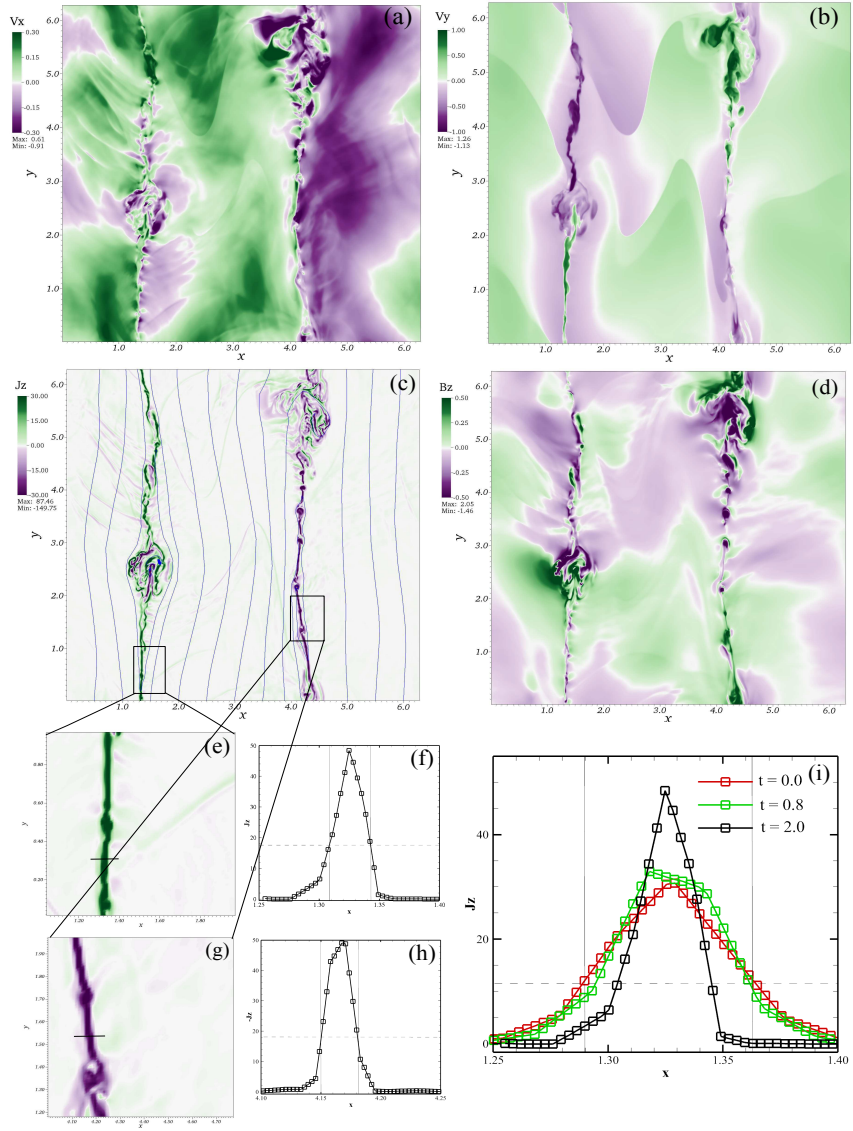}   \\
  \end{tabular}
   \end{center}
\caption{Spatial distributions of a 2D $x-y$ slice at $z = 1.2$ of the velocity components
$V_x$ (panel a) and $V_y$ (panel b), the
current density  $J_z$ component along with the (blue lines) magnetic
field lines projected into the $x-y$ plane (panel c), and the magnetic field component $B_z$
(panel d) for Run A1 at $t = 2.0$.
Panels e and g zoom in
around two strong $|J_z|$ regions; panels f and h show
the one-dimensional (1D) distribution of $|J_z|$ along the black lines in panels e and g;
and panel i shows the 1D distribution of $|J_z|$ at different times, with the black line being the same as that in panel f.
The cell size is $\sim 0.003$ in this run.
}\label{figure2}
\end{figure}

To further demonstrate global reconnection in our simulations, we show in
Figure \ref{figure2} that the classic $X-$point inflow/outflow configuration
is approximately preserved in the turbulent reconnection.
The plasma originating from separate sides of the CSs flows
into the reconnection region with an inflow speed of $\sim 0.15 V_A$,
meanwhile the outflows along the CSs appear to reach values of
$\sim \pm V_A$ (from which the global reconnection rate
could also be estimated to be $\sim 0.15 $ for Run A1).
There seems to be one major
reconnection $X-$point in the left CS near $y=0$
whereas plasmoid-like chains with large $|J_z|$ and $|B_z|$
are visible in the right CS.
In addition, in the left CS  between $y = 0$
and $2$, the strong shear in $|\Delta V_y| / (|B_y|/\sqrt{\rho}) > 2$
might indicate the excitation of Kelvin-Helmholtz
instability \citep{Miura1982, Kowal2020}.

To understand the current sheet structure in more detail, 
we evaluate the current sheet width for Run A1 at different times,
which is defined when $|J_z|$ comes to $e^{-1}$ of
its maximum. In panels f and h of Figure \ref{figure2},
we give two examples of the current sheet width
at $t = 2.0$, in which  the horizontal dashed
lines cut through the $|J_z|$ structure,
and two vertical solid lines mark the current sheet width,
which is about 0.033 and is resolved by about 10 cells.
In panel i, we show the evolution of the current sheet width.
It starts with a width about 0.073 (resolved by about 24 cells),
and undergoes a thinning process but it remains broader than
that predicted by the SP scaling, presumably due to the turbulence.
Overall, the current sheet width is adequately resolved numerically.

\begin{figure}[h!]
     \includegraphics[width = 6. in]{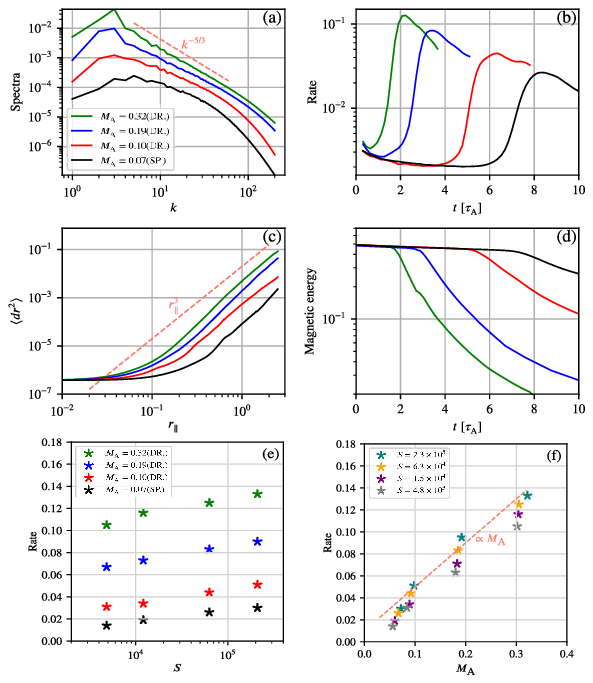}
\caption{Power spectra of kinetic energy
(panel a), global reconnection rate (panel b),
diffusion of reconnecting magnetic field lines (panel c), and magnetic energy evolution
(panel d) for different
levels of turbulence driving for Run A1, B1, C1, and D1, global reconnection rates  as a function
of the Lundquist number $S$ for different values of the Alfv\'en Mach
number $M_{\mathrm{A}}$ (panel e) and as a function
of $M_{\mathrm{A}}$ for different $S$ (panel f).
Quantities in panels (a) and (c) are calculated at the time when
the global reconnection rate shown in panel (b) reaches its
respective maximum. }\label{figure3}
\end{figure}

We now discuss the turbulence properties in further detail.
Panel (a) of
Figure \ref{figure3} shows that the power spectra of kinetic energy
displays a $\sim -5/3$ power law
for different Runs A1 - D1, although the weaker turbulence runs seem to show
slightly flatter spectra.
Because the plasma $\beta \sim 0.1$, the
turbulence is sub-Alfv\'enic but becoming transonic for $M_{\mathrm{A}}\sim 0.3$
(Run A1).
The rate for the strongest external driving turbulence (Run A1) is about
$0.128 \ V_A$, which is consistent with the estimate
measured from inflow and outflow speeds shown in Figure \ref{figure2}.
The rate for spontaneous turbulent reconnection (Run D1) is about
$0.025 \ V_A$, which is basically consistent with the results
by \citet{Oishi2015, Beresnyak2017, Kowal2017}.
In addition, the sharp rise of the rates corresponds well with the
rapid decrease of the total magnetic energy within the simulation box,
indicating that the fast reconnection has dissipated a significant fraction
of the available magnetic energy (by $\sim 50\%$ to nearly $100\%$).

To see the connection between the reconnection rates and
the diffusion of turbulent magnetic fields, we measure the separation
$dr$ of numerous pairs of field lines as a function of $r_\parallel$ (the distance
along the field lines) like \citet{Beresnyak2013}. These pairs start at random positions within
the reconnection regions with $|f_e|<1$.  Panel (c) of Figure \ref{figure3}
shows the relationship between $\langle dr^2\rangle$ and $r_\parallel$ in which $10^5$ field line
pairs are used for statistical averaging.
A stochastic separation of magnetic field lines follows a
super-diffusion behavior. As the reconnection proceeds, $\langle dr^2\rangle$ rises.
At the turbulence injection scale, $r_\parallel \sim 3$, we can calculate the field line separation
rate as $\langle dr^2\rangle^{0.5}/r_\parallel$, similar to \citet{Lazarian1999} and \citet{Eyink2011}.
The rates for Run A1, B1, C1, and D1 are 0.123, 0.093, 0.036, 0.023, respectively,
which are similar to the maximum of the global reconnection rates
obtained from the mixing of traced populations as shown in panel (b).
=However, when $r_\parallel>0.1$,
the standard deviation of the averaged rates for Run A1, B1, C1, and D1 increases to
be the same order of magnitude as the average value $\langle dr^2\rangle$.


Applying these analyses to all the runs (except Run E)
listed in Table 1, we summarize
the dependence of 3D reconnection rates (taken at the peak
of their evolution) on $S$ and $M_{\mathrm{A}}$.
As shown by Panel (e) of Figure \ref{figure3}, the reconnection
rate shows a rather weak
increasing trend as $S$ increases. Even higher $S$ values are needed to see if
the reconnection rate becomes weakly dependent on $S$.
Note that the variation of reconnection rates for a given $M_{\mathrm{A}}$
is within $\sim 50\%$ as $S$ goes from $4.8\times 10^3$ to
$2.3\times 10^5$.
Assuming that the reconnection rate stays below $V_A$ for
very large $S$, it is reasonable to expect that the dependence
of the reconnection rate on $S$ should be rather weak.
Consequently, we conclude that the reconnection rate is weakly dependent on $S$ when $S$ is large.

The reconnection rate, however, does show a clear dependence on the
level of turbulence.
Panel (f) of
Figure \ref{figure3} shows that the rate scales roughly linearly with the turbulent
$M_{\mathrm{A}}$.
This slope is obtained by mostly using points from simulation Runs A-C
with the same $S$.
The weak dependence on $S$ can also be seen.
In addition, it seems that the
``spontaneous" Runs D
cannot be regarded as simply an extrapolation to zero $f_v$, as their
reconnection rates are a bit lower than the extrapolation
from Runs A-C. We suggest that this is due to a fundamental change
of the turbulence properties
between Runs A and Runs D. For Runs A, the turbulence
mostly experiences forward cascades, whereas for Runs D,
the fluctuations are first injected at the CS width scales, then
undergoing both forward and inverse cascades \citep{Bowers2007}.

\begin{figure}[htbp]
     \includegraphics[width = 5. in]{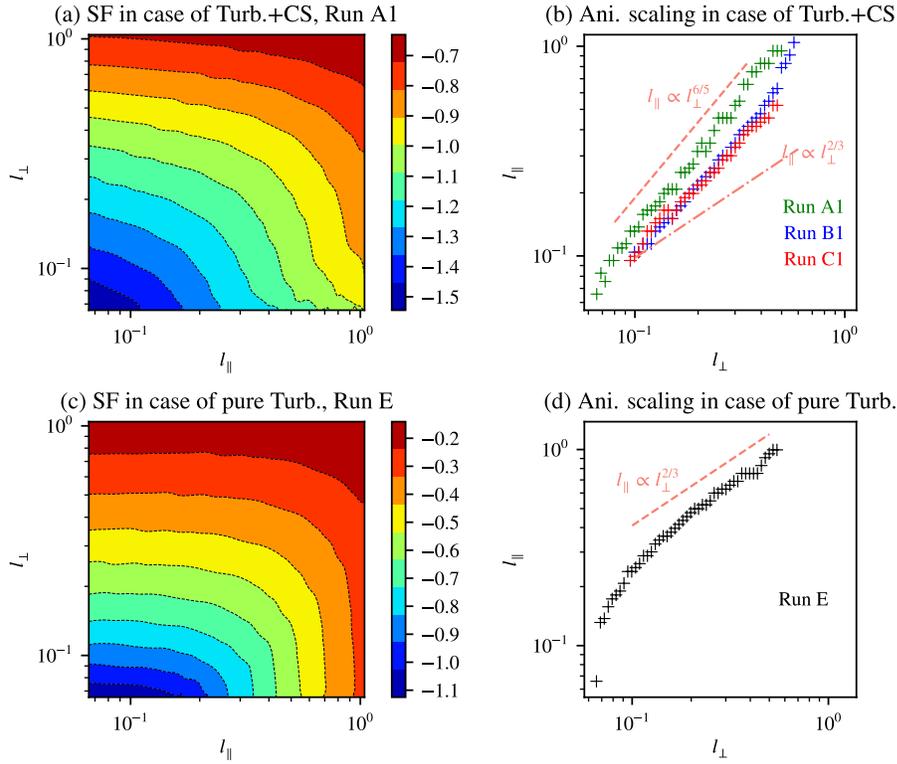}   
\caption{Second-order structure functions (SF) of velocity from
the fully developed turbulence with the large-scale CSs (panel a)
and without the initial large-scale CSs (pure turbulence, panel c); Panels (b) and (d): relationships between semimajor
axis $l_\parallel$ and semiminor axis $l_\perp$ of contours in panels (a) and (c), which measure the scale dependency of turbulent eddy anisotropy. }\label{figure5}
\end{figure}

To investigate the turbulence properties in more detail,
we analyze the anisotropy of the turbulence using Run A1, B1, and C1.
We have calculated the second-order structure functions (SF) of velocity
in terms of parallel  $l_\parallel$  and perpendicular
displacement $l_\perp$ with respect to the local magnetic field
reference frame and the correspondence
between $l_\parallel$  and $l_\perp$ by equating SF values
in parallel and perpendicular directions \citep{Beresnyak2017, Kowal2017}.
The results are shown in Figures \ref{figure5}(a) and (b).
To facilitate a comparison, the results for the fully developed turbulence
without the initial large-scale CSs (Run E) are also presented
in Figures \ref{figure5}(c) and (d).

The resulting SFs clearly display that turbulent eddies
are elongated along the local magnetic field direction for Run A1.
For pure turbulence Run E, eddies become increasingly more
anisotropic at smaller scales, basically conforming to the
Goldreich $\&$ Sridhar prediction \citep{Goldreich1995}.
Comparing the properties from Run A1 and E, however,
we see that the anisotropy in Run A1 is weaker than that in Run E,
showing a power-law scaling
with $l_\parallel \propto l_\perp^{6/5}$
for all the scales captured in the simulation.
Although both Run B1 and Run C1 have smaller Alfv\'en Mach numbers than Run A1,
the anisotropy in them displays a power-law scaling closer to  $l_\parallel \propto l_\perp^{6/5}$ than $l_\parallel \propto l_\perp^{2/3}$.
This may be owing to the fact that how the turbulence is produced in our current models
is different from the traditional \citet{Goldreich1995}'s model,
as discussed in the next section.

\section{DISCUSSION}
\label{sec:discuss}
The results presented here extend the previous studies in 3D turbulent
MHD reconnection by systematically examining the previous unexplored
parameter space in both $S$ and $M_{\mathrm{A}}$.
On the one hand, we find good
consistency with the previous results in the low $S$ and/or low $M_{\mathrm{A}}$
regimes such as the reconnection rates ranging between $0.01 - 0.1 V_A$.
On the other hand, we find two new conclusions: one is that the
reconnection rate is weakly dependent on
$S$ in the large $S$ limit and the other is that the reconnection rate
scales roughly linearly with the turbulent $M_{\mathrm{A}}$.
The weak dependence on $S$ is consistent with both
the turbulent reconnection model \citep{Lazarian1999} and
plasmoid-mediated reconnection model \citep{Loureiro2007,
Bhattacharjee2009, Uzdensky2010, Huang2010}.

The new, linear scaling relationship we find between the reconnection rate
and the strength of turbulence is different from the
 $M_{\mathrm{A}}^2$ scaling given in
\citet{Lazarian1999}. Our turbulence properties are also different
from \citet{Goldreich1995}.
Using the anisotropy scaling from our simulations,
we can derive our new reconnection rate dependence
on $M_{\mathrm{A}}$
in the context of the turbulent reconnection theory
\cite{Lazarian1999, Eyink2011}.
From the constant energy transfer rate of $\dot{\xi} \sim \frac{v_k^2}{\tau_{nl}}  \sim v_k^4 \frac{k_{\perp}^2}{k_{\|} V_\mathrm{A}}$ \citep{Lazarian1999} and the simulation result of $k_{\|} \sim k_{\perp}^{6/5} $, we can get that
$v_{k_{\perp l}} \sim   v_{k_{\perp L}} k_{\perp L}^{1/5}/ k_{\perp l}^{1/5}$, with $L$ being energy injection scale, $l$ being inertial scale, $k_{\perp L} \sim 1/L$, $k_{\perp l} \sim 1/l$, and
$v_{k_{\perp L}}$ as well as  $v_{k_{\perp l}}$ being the corresponding perpendicular fluctuating velocities.
As a pair of field lines with an initial distance of $l_{\perp}^{(0)}$
separate at the rate $\frac{d} {ds} l_{\perp} \sim \frac{ \delta b_{\perp l}} {B_0} \sim \frac{\delta v_{\perp l}} {V_A}$ \citep{Eyink2011}, one finds that
$\frac{d} {ds} l_{\perp} \sim  \frac{v_{k_{\perp L}} k_{\perp L}^{1/5}}{k_{\perp l}^{1/5} V_\mathrm{A}}$, that is $l_{\perp} \sim  M_\mathrm{A}^{5/4} k_{\perp L}^{1/4} s^{5/4}$ with $M_\mathrm{A}=v_{k_{\perp L}}/V_\mathrm{A}$. We can estimate
rate by $l_{\perp}/s \sim  M_A^{5/4}$.
Given that the  inertial range is  limited and the turbulence is not
steady, our numerical result of the rate $\propto M_{\mathrm{A}}$ is
approximately consistent with this relationship.

The nature of turbulence from Runs A to D likely undergoes significant changes.
The turbulence in our simulations come from
both the external driven origin as well as the self-generated origin.
In Run A1, the reconnection rate is high and the flow from the 3D reconnection
is quite significant (the outflow speeds reaching $V_A$ as shown in
Figure \ref{figure2}). Both the presence of large-scale reconnection CSs
and the flows associated with reconnection are affecting the turbulence.
In fact, according to Table 1, comparing Run A2 and Run E where the
external driving $f_v$ is the same (and the same numerical resolution), the turbulent $M_{\mathrm{A}}$
is actually larger in the pure turbulence run
($0.421$) than that in the reconnection run ($0.305$).
Because the self-generated turbulence likely undergoes both forward and
inverse cascades, its spectral properties and anisotropy
will not follow the Goldreich $\&$ Sridhar theory, especially when the turbulence
properties are examined at just a few Alfv\'en times.
In addition, our simulations are in the low $\beta$ situation
(initially at $0.1$) with an aim to model the solar coronal environment
whereas most previous simulations have mostly explored
the higher $\beta$ limit \citep{Oishi2015, Beresnyak2017, Kowal2017}.
According to the results of \citet{Kowal2017},
the anisotropy degree and scaling depend on the plasma $\beta$ and
larger $\beta$ conditions tend to yield scalings closer to
the Goldreich $\&$ Sridhar theory.

The self-generated turbulence/fluctuations likely have several origins.
The first is from the resistive tearing instabilities on relatively smaller
scales of CS thickness; second is from the Kelvin-Helmholtz instability
in the localized outflow regions, again on CS thickness scales; third is from
the ``collisions" of outflows (see $y \sim 2 - 3$ in
Fig. \ref{figure2}). Although these processes can all in principle produce
turbulence, our simulations probably do not have enough
spatial separation to see the development of all these
turbulence. Overall, the third process likely contribute the most to
the self-generated turbulence.

Because the ``spontaneous" Runs can already
produce $M_{\mathrm{A}} \geq 0.06$ with a reconnection rate $\sim 0.01 V_A$,
this implies that, in space and astrophysical systems and to the extent that
periodic boundary conditions can be approximately true,
large-scale current sheets with high $S$
will tend to be destroyed
within several Alfv\'en transient times of the system.



\begin{acknowledgments}

This work is supported by NSFC grants under contracts 41974171, 41774157, 41731067 and 41674171.
HL, FG, XL and SL acknowledge support by LANL/LDRD program and DoE/OFES program.
FGs contributions are in part based upon work supported by the U.S. Department of Energy, Office of Fusion Energy Science, under Award Number de-sc0018240 and by NSF grant AST-1735414.
Institutional Computing resources at LANL and resources of the National Energy Research Scientific Computing Center (NERSC) are used for simulations reported here.

\end{acknowledgments}

\bibliographystyle{apj}

\end{document}